\documentclass[runningheads]{llncs}
\usepackage[T1]{fontenc}
\usepackage[utf8]{inputenc}
\usepackage{amssymb} 
\usepackage{amsmath} 
\allowdisplaybreaks
\usepackage{subcaption}
\usepackage{defs}
\graphicspath{{./img/}}

\newcommand{\numUniqueClasses}{104}

\newcommand{\EffSizetypehintsapimdA}{1.0}

\newcommand{\EffSizetypehintsasyncbtreeA}{0.5345604808414726}

\newcommand{\EffSizetypehintscodetimingA}{0.6362795857988166}

\newcommand{\EffSizetypehintscodetimingPvalue}{0.005313020883379603}

\newcommand{\EffSizetypehintsmimesisA}{0.5298047899306567}

\newcommand{\EffSizetypehintspythonstringutilsA}{0.704952741020794}

\newcommand{\AvgEffSizetypehints}{0.6177798788654812}

\newcommand{\AvgEffSizenotypes}{0.6025370325680607}

\newcommand{\AvgEffSizews}{0.5778231398547163}

\newcommand{\EffSizerandompyparaA}{0.41058787319175133}

\newcommand{\EffSizerandompyparaPvalue}{0.009371511338051804}

\newcommand{\AvgEffSizerandom}{0.5544051467139133}

\begin{document}

\title{Automated Unit Test Generation for Python %
}

\author{Stephan Lukasczyk \and
Florian Kroiß \and
Gordon Fraser}

\authorrunning{S. Lukasczyk et al.}

\institute{University of Passau, Innstr. 33, 94032 Passau, Germany
\email{\{stephan.lukasczyk,gordon.fraser\}@uni-passau.de,
kroiss@fim.uni-passau.de}}
\maketitle              
\begin{abstract}
  %
  Automated unit test generation is an established research field, and mature
  test generation tools exist for statically typed programming languages such
  as Java.
  %
  It is, however, substantially more difficult to automatically generate
  supportive tests for dynamically typed programming languages such as Python,
  due to the lack of type information and the dynamic nature of the language.
  %
  In this paper, we describe a foray into the problem of unit test generation
  for dynamically typed languages.
  %
  We introduce \pynguin, an automated unit test generation framework for
  Python. Using \pynguin, we aim to empirically shed light on two central
  questions: (1) Do well-established search-based test generation methods,
  previously evaluated only on statically typed languages, generalise to
  dynamically typed languages? (2) What is the influence of incomplete type
  information and dynamic typing on the problem of automated test generation?
  %
  Our experiments confirm that evolutionary algorithms can outperform random
  test generation also in the context of Python, and can even alleviate the
  problem of absent type information to some degree. However, our results
  demonstrate that dynamic typing nevertheless poses a fundamental issue for
  test generation, suggesting future work on integrating type inference.
\end{abstract}

\section{Introduction}\label{sec:introduction}

Unit tests can be automatically generated to support developers and the dynamic
analysis of programs. Established techniques such as feedback-directed random
test generation~\cite{PLE+07} or evolutionary algorithms~\cite{FA11} are
implemented in mature research prototypes, but these are based on strong
assumptions on the availability of static type information, as is the case in
statically typed languages like Java. Dynamically typed languages such
as Python or JavaScript, however, have seen increased popularity within recent years.
%
%
Python is the most popular programming language
in the category of dynamically typed languages,
according to,
for example,
the IEEE Spectrum Ranking\footnote{%
  \url{https://spectrum.ieee.org/computing/software/the-top-programming-languages-2019},
  accessed 2020–07–25.
}.
It is heavily used in the fields of machine learning and data analysis, and
it is also popular in other domains.
This can be seen,
for example,
from the Python Package Index~(PyPI),
which contains more than \num{200000} packages at the time of writing.
In languages like Python, the type information that automated unit test generators require is not available.

An automated unit test generator primarily requires type information in order
to select parameters for function calls and to generate complex objects. If
type information is absent, the test generator can only guess which calls to
use to create new objects, or which existing objects to select as parameters
for new function calls.
Existing test generators for dynamically typed languages therefore resort to other means to avoid having to make such choices in the first place, for example by using the document object model of a web browser
to generate tests for JavaScript~\cite{MMP15}, or by targetting the browser's event handling system rather than APIs~\cite{ADJ+2011,LAG14}. However, there is no general purpose unit test generator at API level yet for languages like Python.

In order to allow test generation research to expand its focus from statically to dynamically typed languages, in this paper we introduce \pynguin, a new automated test generation framework for Python.
\pynguin takes as input a Python module and its dependencies, and aims to
automatically generate unit tests that maximise code coverage. In order to
achieve this, \pynguin implements the established test generation techniques of
whole-suite generation~\cite{FA13} and feedback-directed random
generation~\cite{PLE+07}.
\pynguin is available as open source 
to support future research on automated test generation for
dynamically typed programming languages.
\pynguin is designed to be extensible; in this paper we focus on established
baseline
algorithms for foundational experiments, we will add further
algorithms such as DynaMOSA~\cite{PKT18} in future work.
Using \pynguin, we empirically study the problem of automated unit test
generation for Python using ten popular open source Python projects taken from
GitHub, all of which contain type information added by developers in terms of
type annotations. This selection allows us to study two central questions:
(1)~Do previous findings, showing that evolutionary search achieves higher code
coverage than random testing~\cite{campos2018empirical}, also generalise to
dynamically typed languages? (2)~What is the influence of the lack of type
information in a dynamically typed language like Python on automated unit test
generation?
In detail, the contributions of this paper are the following:
\begin{enumerate}
  \item We introduce \pynguin, a new framework for automated unit test
  generation for the Python programming language.
  \item We replicate experiments previously conducted only in the context of statically typed languages to compare test generation approaches.
  \item We empirically study the influence of type information on the effectiveness of automated test generation.
\end{enumerate}

Our experiments confirm that the whole-suite approach generally achieves higher
code coverage than random testing, and that the availability of type
information also leads to higher resulting coverage. However, our experiments
reveal several new technical challenges 
such as generating collections or iterable input types.
Our findings also suggest that the integration of current research on type
inference is a promising route forward for future research.

\section{Background}\label{sec:background}

The main approaches to automatically generate unit tests are either by creating random sequences, or by applying metaheuristic search algorithms.
Random testing assembles sequences of calls to constructors and methods randomly, often with the objective to find undeclared exceptions~\cite{csallner2004jcrasher} or violations of general object contracts~\cite{PLE+07}, but the generated tests can also be used as automated regression tests. The effectiveness of random test generators can be increased by integrating heuristics~\cite{ma2015grt,sakti2014instance}. Search-based approaches use a similar representation, but apply evolutionary search algorithms to maximize code coverage~\cite{tonella2004evolutionary,baresi2010testful,FA13,andrews2011genetic}.

As an example to illustrate how type information is used by existing test
generators, consider the following snippets
of Java (left) and Python (right) code:
\begin{minipage}[t]{0.48\textwidth}
  \begin{lstlisting}[language=Java]
class Foo {
  Foo(Bar b) { ... }
  void doFoo(Bar b) { ... } }
class Bar {
  Bar() { ... }
  Bar doBar(Bar b) { ... } }
  \end{lstlisting}
\end{minipage}%
\hfill%
\begin{minipage}[t]{0.48\textwidth}
  \begin{lstlisting}[language=Python]
class Foo:
  def __init__(self, b): ...
  def do_foo(self, b): ...
class Bar:
  def __init__(self): ...
  def do_bar(self, b): ...
  \end{lstlisting}
\end{minipage}

Assume \mintinline{java}{Foo}
of the Java example
is the class under test. It has a dependency on
class \mintinline{java}{Bar}: in order to generate an object of type
\mintinline{java}{Foo} we need an instance of \mintinline{java}{Bar}, and the
method \mintinline{java}{doFoo} also requires a parameter of type
\mintinline{java}{Bar}.

Random test generation would typically generate tests in a forward way.
Starting with an empty sequence $t_0 = \langle \rangle$, all available calls
for which all parameters can be satisfied with objects already existing in the
sequence can be selected. In our example, initially only the constructor of
\mintinline{java}{Bar} can be called, since all other methods and constructors
require a parameter, resulting in $t_1 = \langle o_1 = $ \mintinline{java}{new
Bar()}$\rangle$. Since
$t_1$ contains an object of type \mintinline{java}{Bar}, in the second step the
test generator now has a choice of either invoking \mintinline{java}{doBar} on
that object (and use the same object also as parameter), or invoking the
constructor of \mintinline{java}{Foo}. Assuming the chosen call is the
constructor of \mintinline{java}{Foo}, we now have $t_2 = \langle o_1 = $
\mintinline{java}{new Bar()}$; o_2 = $ \mintinline{java}{new
Foo(}$o_1$\mintinline{java}{)}$; \rangle$. Since there now is also an instance
of
\mintinline{java}{Foo} in the sequence, in the next step also the method
\mintinline{java}{doFoo} is an option. The random test generator will continue
extending the sequence in this manner, possibly integrating heuristics to
select more relevant calls, or to decide when to start with a new sequence, etc.

An alternative approach, for example applied during the mutation step of an
evolutionary test generator, is to select necessary calls in a backwards
fashion. That is, a search-based test generator like \evosuite~\cite{FA13}
would first decide that it needs to, for example, call method
\mintinline{java}{doFoo} of class \mintinline{java}{Foo}. In order to achieve
this, it requires an instance of \mintinline{java}{Foo} and an instance of
\mintinline{java}{Bar} to satisfy the dependencies. To generate a parameter
object of type \mintinline{java}{Bar}, the test generator would consider all
calls that are declared to return an instance of
\mintinline{java}{Bar}---which is the case for the constructor of
\mintinline{java}{Bar} in our example, so it would prepend a call to \mintinline{java}{Bar()}
before the invocation of \mintinline{java}{doFoo}. Furthermore, it would try to
instantiate \mintinline{java}{Foo} by calling the constructor. This, in turn,
requires an instance of \mintinline{java}{Bar}, for which the test generator
might use the existing instance, or could invoke the constructor of
\mintinline{java}{Bar}.

In both scenarios, type information is crucial: In the forward construction
type information is used to inform the choice of call to append to the
sequence, while in the backward construction type information is used to select
generators of dependency objects.
Without type information,
which is the case with the Python example,
a forward construction (1) has to allow all possible functions at all steps,
thus may not only select the constructor of \mintinline{python}{Bar},
but also that of \mintinline{python}{Foo} with an arbitrary parameter type,
and (2) has to consider all existing objects for all parameters of a selected
call, and thus, for example, also \mintinline{python}{str} or \mintinline{python}{int}.
Backwards construction without type information would also have to try to
select generators from all possible calls, and all possible objects,
which both result in a potentially large search space to select from.
Type information can be provided in two ways in recent Python versions:
either in a stub file that contains type hints
or directly annotated in the source code.
A stub file can be compared to C header files:
they contain,
for example,
method declarations with their according types.
Since Python~3.5,
the types can also be annotated directly in the implementing source code,
in a similar fashion known from statically typed languages~(see
PEP~484\footnote{%
  \url{https://python.org/dev/peps/pep-0484/}, accessed 2020–07–25.
}).

\section{Search-based Unit Test Generation}\label{sec:approach}

\subsection{Python Test Generation as a Search Problem}

As the \emph{unit} for unit test generation, we consider Python \emph{modules}.
A module is usually identical with a file
and contains definitions of,
for example,
functions, classes, or statements;
these can be nested almost arbitrarily.
When the module is loaded
the definitions and statements at the top level are executed.
While generating tests
we do not only want all definitions to be executed,
but also all structures defined by those definitions,
for example,
functions, closures, or list comprehensions.
Thus,
in order to apply a search algorithm,
we first need to define a proper representation of the valid solutions
for this problem.

We use a representation
based on prior work from the domain of testing Java code~\cite{FA13}.
For each statement~\(s_j\) in a test case~\(t_i\)
we assign one value~\(v(s_j)\)
with type~\(\tau(v(s_j)) \in \mathcal{T}\),
with the finite set of types~\(\mathcal{T}\) used in the
subject-under-test~(SUT)
and the modules imported by the SUT.
%
We define four kinds of statements:
\emph{Primitive statements} represent \mintinline{python}{int},
\mintinline{python}{float}, \mintinline{python}{bool},
and \mintinline{python}{str} variables,
for example, \mintinline{python}{var0 = 42}.
Value and type of a statement are defined by the primitive variable.
Note that although in Python everything is an object,
we treat these values as primitives
because they do not require further construction in Python's syntax.
Other simple types,
such as lists,
require the construction of the list and its elements,
which we do not yet handle.
%
\emph{Constructor statements} create new instances of a class,
for example, \mintinline{python}{var0 = SomeType()}.
Value and type are defined by the constructed object;
any parameters are satisfied from the set~\(V = \{v(s_k) \mid 0 \leq k < j\}\).
\emph{Method statements} invoke methods on objects,
for example, \mintinline{python}{var1 = var0.foo()}.
Value and type are defined by the return value of the method;
source object and any parameters are satisfied from the set~\(V\)\kern-0.2em .
\emph{Function statements} invoke functions,
for example, \mintinline{python}{var2 = bar()}.
They do not require a source object
but are otherwise identical to method statements.
This representation is of variable size;
we constrain the size of test cases~\(l\in[1, L]\)
and test suites~\(n\in[1, N]\).
%
In contrast to prior work on testing Java~\cite{FA13},
we do not define field or assignment statements;
fields of objects are not explicitly declared in Python
but assigned dynamically,
hence it is non-trivial to identify the existing fields of an object
and we leave it as future work.
%

The search operators for this representation are based on those used in \evosuite~\cite{FA13}:
Crossover takes as input two test suites~\(P_1\) and~\(P_2\),
and generates two offspring~\(O_1\) and~\(O_2\).
Individual test cases have no dependencies between each other,
thus the application of crossover
always generates valid test suites as offspring.
Furthermore,
the operator decreases the difference in the number of test cases
between the test suites,
thus \(\abs (|O_1|-|O_2|) \leq \abs(|P_1|-|P_2|)\).
Therefore,
no offspring will have more test cases than the larger of its parents.

When mutating a test suite~\(T\),
each of its test cases is mutated with probability~\(\frac{1}{|T|}\).
After mutation,
we add new randomly generated test cases to~\(T\).
The first new test case is added with probability~\(\sigma\).
If it is added,
a second new test case is added with probability~\(\sigma^2\);
this happens until the \(i\)-th test case
is not added~(probability: \(1-\sigma^i\)).
Test cases are only added
if the limit~\(N\) has not been reached,
thus~\(|T| \leq N\).
The mutation of a test case
can be one of three operations:
\emph{remove}, which removes a statement from the test case,
\emph{change}, which randomly changes values in a statement—for example, by
adding random values to numbers, adding/replacing/deleting
characters, or changing method calls—and \emph{insert}, which adds new
statements at random positions
in the test case.
Each of these operations can happen with the same probability
of~\(\frac{1}{3}\).
A test case that has no statements left after the application
of the mutation operator
is removed from the test suite~\(T\).
For constructing the initial population, a random test case~\(t\)
is sampled by uniformly choosing a value~\(r\) with \(1 \leq r \leq L\),
 and then applying the insertion operator repeatedly starting with an empty test case~\(t'\),
until~\(|t'| \geq r\).

\subsection{Covering Python Code}\label{sec:approach-covering}

A Python module contains various control structures,
for example,
\mintinline{python}{if} or \mintinline{python}{while} statements,
which are guarded by logical predicates.
The control structures are represented by conditional jumps
at the bytecode level,
based on either a unary or binary predicate.
We focus on \emph{branch coverage} in this work,
which requires that each of those predicates evaluates to
both true and false.
Let~\(B\) denote the set of branches in the SUT---two
for each conditional jump in the byte code.
Everything executable in Python is represented as a \emph{code object}.
For example,
an entire module is represented as a code object,
a function within that module is represented as another code object.
We want to execute all code objects~\(C\) of the SUT.
Therefore,
we keep track of the executed code objects~\(C_T\)
as well as the minimum \emph{branch distance}~\(d_{\min}(b, T)\)
for each branch~\(b \in B\),
when executing a test suite~\(T\).
\(B_T \subseteq B\) denotes the set of taken branches.
We then define the branch coverage~\(\cov(T)\) of a test suite~\(T\) as
\(  \cov(T) = \frac{|C_T| + |B_T|}{|C| + |B|} \).

The fitness function required by the genetic algorithm of our whole-suite approach
is constructed similar to the one used in \toolname{EvoSuite}~\cite{FA13}
by incorporating the branch distance.
\emph{Branch distance} is a heuristic to determine
how far a predicate is away from evaluating to true or false, respectively.
In contrast to previous work on Java,
where most predicates at the bytecode level operate only
on Boolean or numeric values,
in our case the operands of a predicate can be any Python object.
Thus,
as noted by Arcuri~\cite{A13},
we have to define our branch distance in such a way
that it can handle arbitrary Python objects.
Let~\(\mathbb{O}\) be the set of possible Python objects
and let~\(\Theta := \{\equiv, \not\equiv, <, \leq, >, \geq, \in, \notin,
=, \neq\}\) be the set of binary comparison operators~(remark: we use
\enquote{\(\equiv\)}, \enquote{\(=\)}, and \enquote{\(\in\)} for Python's
\mintinline{python}{==}, \mintinline{python}{is}, and \mintinline{python}{in}
keywords, respectively).
For each~\(\theta \in \Theta\),
we define a function~\(\delta_\theta: \mathbb{O}\times\mathbb{O} \to
\mathbb{R}_0^+ \cup \{\infty\}\)
that computes the branch distance of the true branch of a predicate of the form
\(a \mathbin{\theta} b\), with \(a,b\in\mathbb{O}\) and \(\theta\in\Theta\).
By~\(\delta_{\bar \theta}(a,b)\) we denote the distance of the false branch,
where~\(\bar \theta\) is the complementary operator of~\(\theta\).
Let further \(k\) be a positive number,
and let~\(\lev(x,y)\) denote the Levenshtein distance~\cite{L66}
between two strings~\(x\) and~\(y\).
The predicates \(\mathrm{is\_numeric}(z)\) and \(\mathrm{is\_string}(z)\)
determine whether the type of their argument~\(z\) is numeric or a string,
respectively.
\begin{align*}
\delta_{\equiv}(a, b) &= \begin{cases}
0 & a \equiv b \\
|a-b| & a \not\equiv b \land \text{is\_numeric}(a) \land \text{is\_numeric}(b) \\
\text{lev}(a,b) & a \not\equiv b \land \text{is\_string}(a) \land \text{is\_string}(b) \\
\infty & \text{otherwise}
\end{cases}
\\
\delta_{<}(a, b) &= \begin{cases}
0 & a < b \\
a-b+k & a \geq b \land \text{is\_numeric}(a) \land \text{is\_numeric}(b) \\
\infty & \text{otherwise}
\end{cases}
\\
\delta_{\leq}(a, b) &= \begin{cases}
0 & a \leq b \\
a-b+k & a > b \land \text{is\_numeric}(a) \land \text{is\_numeric}(b) \\
\infty & \text{otherwise}
\end{cases}
\\
\delta_{>}(a, b) &= \delta_{<}(b,a)
\\
\delta_{\geq}(a, b) &= \delta_{\leq}(b,a)
\\
\delta_{\theta}(a, b) &= \begin{cases}
0 & a \mathbin{\theta} b \\
k & \text{otherwise}
\end{cases}
\qquad \theta \in \{\not\equiv, \in, \notin, =, \neq\}
\end{align*}
Note that every object in Python represents a Boolean value
and can therefore be used as a predicate.
We assign a distance of~\(0\) to the true branch
of such a unary predicate,
if the object represents a true value,
otherwise~\(k\).
Future work shall refine the branch distance for different operators
and operand types.
The fitness function estimates how close a test suite is
to covering \emph{all} branches of the SUT.
Thus,
every predicate has to be executed at least twice,
which we enforce in the same way as existing work~\cite{FA13}:
the actual branch distance~\(d(b, T)\) is given by
\begin{align*}
d(b,T) &= \begin{cases}
0 & \text{if the branch has been covered}\\
\nu(d_{min}(b,T)) & \text{if the predicate has been executed at least twice}\\
1 & \text{otherwise}
\end{cases}
\end{align*}
with \(\nu(x) = \frac{x}{x+1}\) being a normalisation function~\cite{FA13}.
Finally,
we can define the resulting fitness function~\(f\) of a test suite~\(T\) as
\begin{align*}
  f(T) &= |C| - |C_T| + \sum_{b \in B} d(b, T)
\end{align*}

\subsection{The \pynguin{} Framework}\label{sec:approach-pinguin}

\pynguin{} is a framework for automated unit test generation
written in and for the Python programming language.
The framework is available as open-source software
licensed under the GNU Lesser General Public License
from its GitHub repository\footnote{%
  \url{https://github.com/se2p/pynguin},
  accessed 2020–07–27.
}.
It can also be installed from the Python Package Index~(PyPI)\footnote{%
  \url{https://pypi.org/project/pynguin/},
  accessed 2020–07–25.
} using the \command{pip} utility.
\pynguin takes as input a Python module
and allows the generation of unit tests
using different techniques.
For this, it parses the module
and extracts information about available methods in the module
and types from the module and its imports.
%
So far,
\pynguin focuses on test-input generation
and excludes the generation of oracles.
A tool run emits the generated test cases
in the style of the widely-used \toolname{PyTest}\footnote{%
  \url{https://www.pytest.org}, accessed 2020–07–25.%
} framework
or for the \texttt{unittest} module from the Python standard library.

\pynguin is built to be extensible with other test generation approaches and algorithms. For experiments in this paper, we implemented a feedback-directed random approach based on \toolname{Randoop}~\cite{PLE+07} in addition to the whole-suite test-generation approach.
Feedback-directed test generation starts with two empty test suites,
a passing and a failing test suite,
and adds statements randomly to an empty test case.
After each addition,
the test case is executed
and the execution result is retrieved.
Successful test cases,
that is,
test cases that do not raise exceptions
are added to the passing test suite;
a test case that raises an exception
is added to the failing test suite.
In the following,
the algorithm randomly chooses a test case from the passing test suite
or an empty test case
and adds statements to it.
We refer the reader to the description of \toolname{Randoop}~\cite{PLE+07}
for details on the algorithm;
the main differences of our approach are
that it does not yet check for contract violations,
and does not require the user to provide a list of relevant classes and methods,
which \toolname{Randoop} does.
%

\section{Experimental Evaluation}\label{sec:evaluation}

Using our \pynguin test generator, we aim to empirically study automated unit test generation on Python. First, we are interested in determining whether previous findings on the performance of test generation techniques established in the context of statically typed languages generalise also to Python:
\begin{resq}[RQ\ref{rq:quality}]\label{rq:quality}
  How do whole-suite test generation and random test generation compare on Python code?
\end{resq}

A central difference between prior work and the context of Python is the type
information: Previous work evaluated test-generation techniques mainly for
statically typed languages, such as Java, where information on parameter types
is available at compile time, that is, without running the program.
This is not the case for many programs written in dynamically typed languages,
such as Python.
Therefore, we want to explicitly evaluate the influence of the type information
for the test-generation process:
\begin{resq}[RQ\ref{rq:typeinfluence}]\label{rq:typeinfluence}
  How does the availability of type information influence test generation?
\end{resq}

\subsection{Experimental Setup}

In order to answer the two research questions,
we created a dataset of Python projects for experimentation.
We used the \enquote{typed} category
of the PyPI package index of Python projects,
and selected ten projects by searching for projects
that contain type hints in their method signatures,
and that do not have dependencies to native-code libraries,
such as \texttt{numpy}.
Details of the chosen projects are shown in Table~\ref{tab:projects}:
the column \emph{Project Name} gives the name of the project on PyPI;
the lines of code were measured with the \toolname{cloc}\footnote{%
  \url{https://github.com/AlDanial/cloc}, accessed 2020–07–25.%
} utility tool.
The table furthermore shows the absolute average number of code objects,
predicates,
and detected types per module of each project.
The former two measures give an insight on the project's complexity;
higher numbers indicate larger complexity.
The latter provides an overview how many types \pynguin was able to parse~(note
that \pynguin may not be able to resolve all types).
\begin{table}[t!]
  \caption{\label{tab:projects}Projects used for evaluation}
  \centering
  \begin{tabular}{
      @{}
      l
      r
      S[table-number-alignment=right,round-precision=4]
      S[table-number-alignment=right]
      S[table-number-alignment=right,round-mode=places,round-precision=1]
      S[table-number-alignment=right,round-mode=places,round-precision=1]
      S[table-number-alignment=right,round-mode=places,round-precision=1]
      @{}
    } \toprule
    Project Name
    & Version
    & \multicolumn{1}{c}{~~~~LOCs~~~}
    & \multicolumn{1}{r}{Modules~~}
    & \multicolumn{1}{r}{CodeObjs.~~}
    & \multicolumn{1}{r}{Preds.~~}
    & \multicolumn{1}{r}{Types} \\ \midrule
    %
    %
    \verb|apimd|&1.0.2&316&1&35.0&83.0&11.0\\%
    \verb|async_btree|&1.0.1&284&6&9.0&8.666666666666666&6.333333333333333\\%
    \verb|codetiming|&1.2.0&85&2&18.0&8.0&6.0\\%
    \verb|docstring_parser|&0.7.1&608&6&12.0&15.666666666666666&9.5\\%
    \verb|flutes|&0.2.0.post0&1085&9&19.0&26.0&5.0\\%
    \verb|flutils|&0.6&1715&13&10.23076923076923&22.307692307692307&8.384615384615385\\%
    \verb|mimesis|&4.0.0&1663&34&12.333333333333334&5.7&9.166666666666666\\%
    \verb|pypara|&0.0.22&1305&6&47.166666666666664&23.5&12.0\\%
    \verb|python-string-utils|&1.0.0&476&4&21.0&29.5&6.5\\%
    \verb|pytutils|&0.4.1&1108&23&8.1875&6.625&6.0625\\%
    \midrule%
    Total&&8645&104&191.91826923076923&228.96602564102562&79.94711538461539\\
    \bottomrule
  \end{tabular}
\end{table}
The central metric we use to evaluate the performance of a test generation technique is code coverage. In particular, we measure branch coverage at the level of bytecode; like in Java bytecode, complex conditions are compiled to nested branches with atomic conditions also in Python code. In addition to the final overall coverage, we also keep track of coverage over time to shed light on the speed of convergence.
In order to statistically compare results we use the Mann-Whitney U-test and
the Vargha and Delaney effect size~\effectsize.

We executed \pynguin in four different configurations: First, we executed \pynguin using random test generation and whole test suite generation; second, we ran \pynguin with the developer-written type annotations contained in the projects, and without them. To answer RQ1, we compare the performance of random test generation and whole test suite generation; to answer RQ2 we compare the performance of each of these techniques for the case with and without type information.

For each project, \pynguin was run on each of the constituent modules in sequence. We executed \pynguin{} in \toolname{git} revision 5f538833 in a Docker container that is based on Debian~10 and utilises Python~3.8.3.
In line with previous work, we set the maximum time limit for the test-generation algorithms,
that is,
the time without analysing the module-under-test
and without export of the results,
 to \SI{600}{\second} per module.
%
%
We ran \pynguin \num{30} times on each module and configuration
to minimise the influence of randomness.
All experiments were executed on dedicated compute servers equipped with
Intel Xeon E5-2690v2 CPUs and \SI{64}{\giga\byte} RAM, running Debian~10.
All scripts and the raw data are available as supplementary material\footnote{%
  \url{https://github.com/se2p/artifact-pynguin-ssbse2020}, accessed 2020–07–27.
}.

\subsection{Threats to Validity}\label{sec:evaluation-threats}

\paragraph{Internal Validity}
The standard coverage tool for Python is \toolname{Coverage.py},
which offers the capability to measure branch coverage.
However, it measures branch coverage by comparing which transitions between source lines have occurred and which are possible.
This method of measuring branch coverage is imprecise,
because not every branching statement necessarily leads to a source line transition,
for example, \mintinline{Python}{x = 0 if y > 42 else 1337}.
We thus implemented our own coverage measurement.
We tried to mitigate possible errors in our implementation,
by providing sufficient unit tests for it.

\paragraph{External Validity}
We used \num{\numUniqueClasses} modules
from ten different Python projects
for our experiments. It is conceivable that the exclusion of projects without
type annotations or native-code libraries leads to a selection of smaller
projects, and the results may thus not generalise to other Python projects.
However, besides the two constraints listed, no others were applied during the
selection.

\paragraph{Construct Validity}

Methods called with wrong input types may still cover parts of the code before possibly raising exceptions due to the invalid inputs.
We conservatively included all coverage in our analysis, which may improve
coverage for configurations that ignore type information, and thus reduce the
effect we observed. However, it does not affect our general conclusions.
Further, we cannot measure fault finding capability as our tool does not
generate assertions, which is explicitly out of scope of this work.

\subsection{RQ1: Whole-suite Test Generation vs. Random Testing}\label{sec:rq1}

\begin{figure}[t!]
  \begin{subfigure}[t]{0.485\textwidth}
    \centering
    \includegraphics[width=\textwidth]{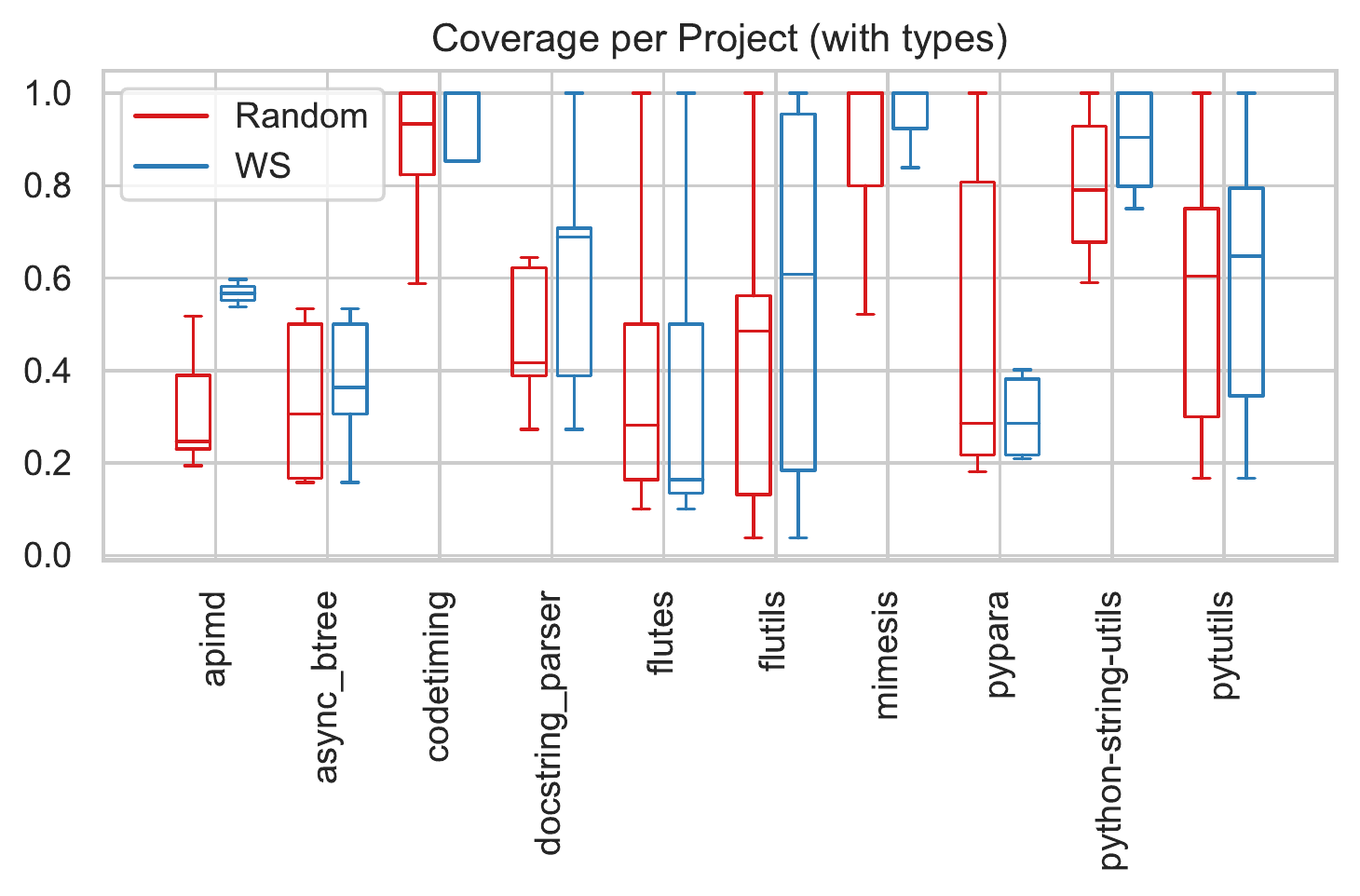}
    \caption{\label{fig:coverage-per-project-typehints}With type information}
  \end{subfigure}%
  \hfill%
  \begin{subfigure}[t]{0.485\textwidth}
    \centering
    \includegraphics[width=\textwidth]{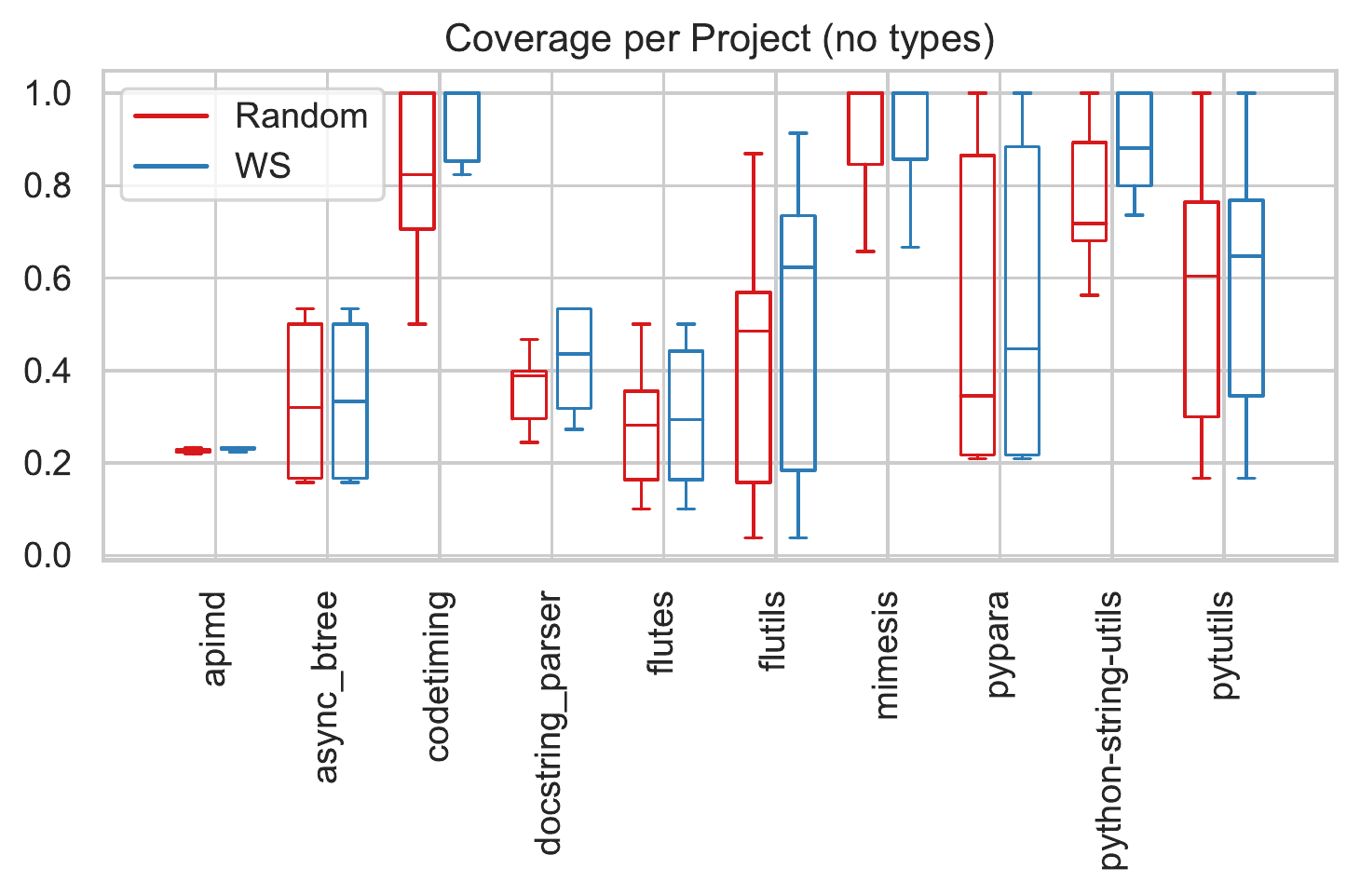}
    \caption{\label{fig:coverage-per-project-notypes}Without type information}
  \end{subfigure}
  \caption{\label{fig:coverage-per-project}Coverage per project and
  configuration}
\end{figure}

Figure~\ref{fig:coverage-per-project} provides an overview
over the achieved coverage per project in box plots.
Each data point in the plot is one achieved coverage value for one of the modules of the project.
Figure~\ref{fig:coverage-per-project-typehints} reports the coverage values
for whole-suite and random test generation with available type hints,
whereas Fig.~\ref{fig:coverage-per-project-notypes} reports the same
without the usage of type hints to guide the generation.
Coverage values
range from \SIrange{0}{100}{\percent}
depending on the project.
The coverage achieved varies between projects, with some projects achieving
generally high coverage (for example, \texttt{python-string-utils},
\texttt{mimesis}, \texttt{codetiming}), and others posing challenges for
\pynguin (for example,
\texttt{apmid}, \texttt{async\_btree}, \texttt{pypara}, \texttt{flutes}). For
example, for the \texttt{apimd} project without type information the coverage
is slightly above \SI{20}{\percent}, which is the coverage achieved just by importing the module.
In Python,
when a module is imported,
the import statements of the module,
as well as class and method definitions
are executed,
and thus covered.
Note that this does not execute the method bodies.
For other projects with low coverage,
\pynguin is not able to generate reasonable inputs,
for example,
higher-order functions or collections,
due to technical limitations.

\begin{figure}[t!]
  \begin{subfigure}[t]{0.475\textwidth}
    \centering
    \includegraphics[width=\textwidth]{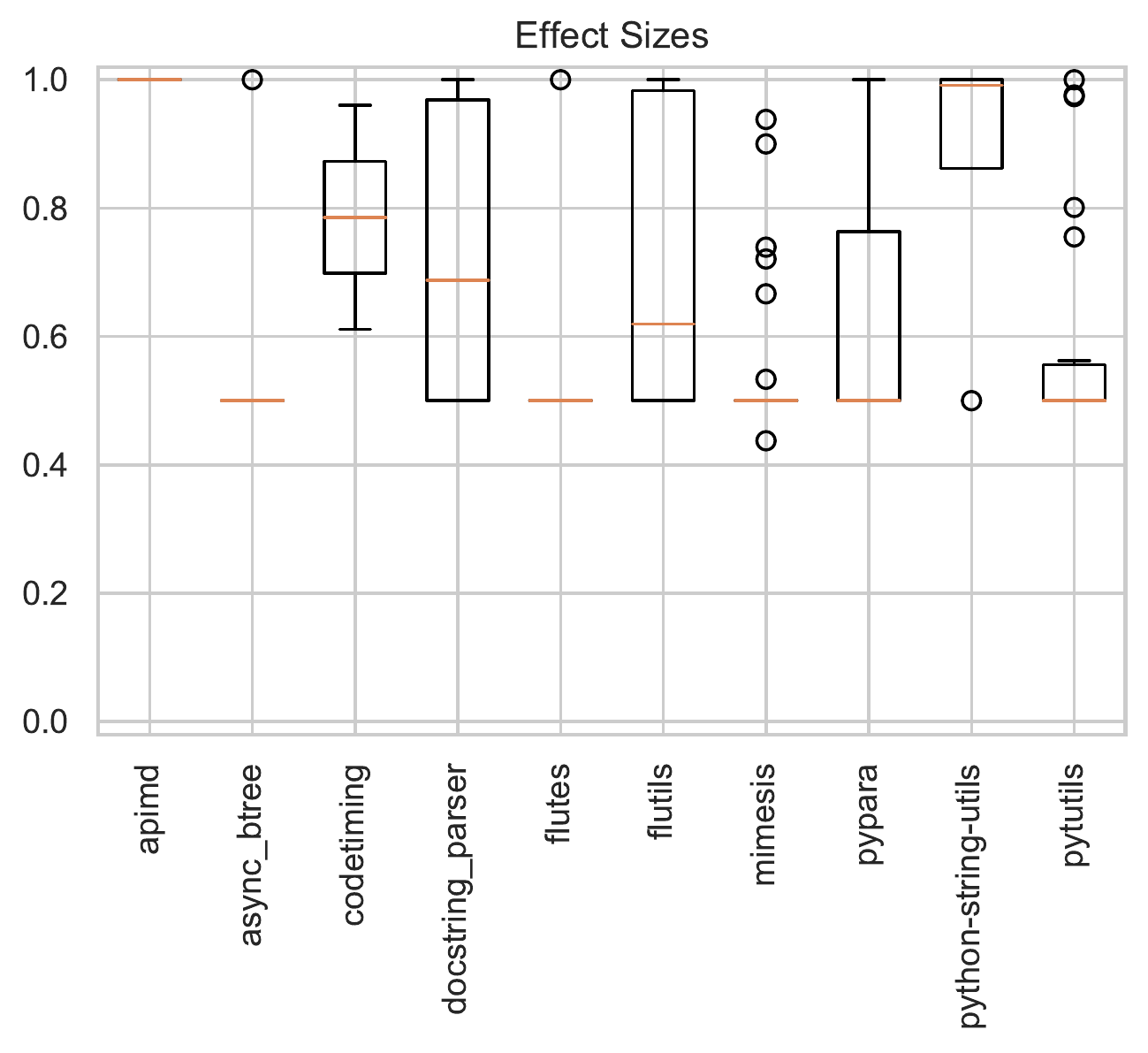}
    \caption{\label{fig:effect-sizes-typehints}\effectsize~effect sizes with
      type information}
  \end{subfigure}%
  \hfill%
  \begin{subfigure}[t]{0.475\textwidth}
    \centering
    \includegraphics[width=\textwidth]{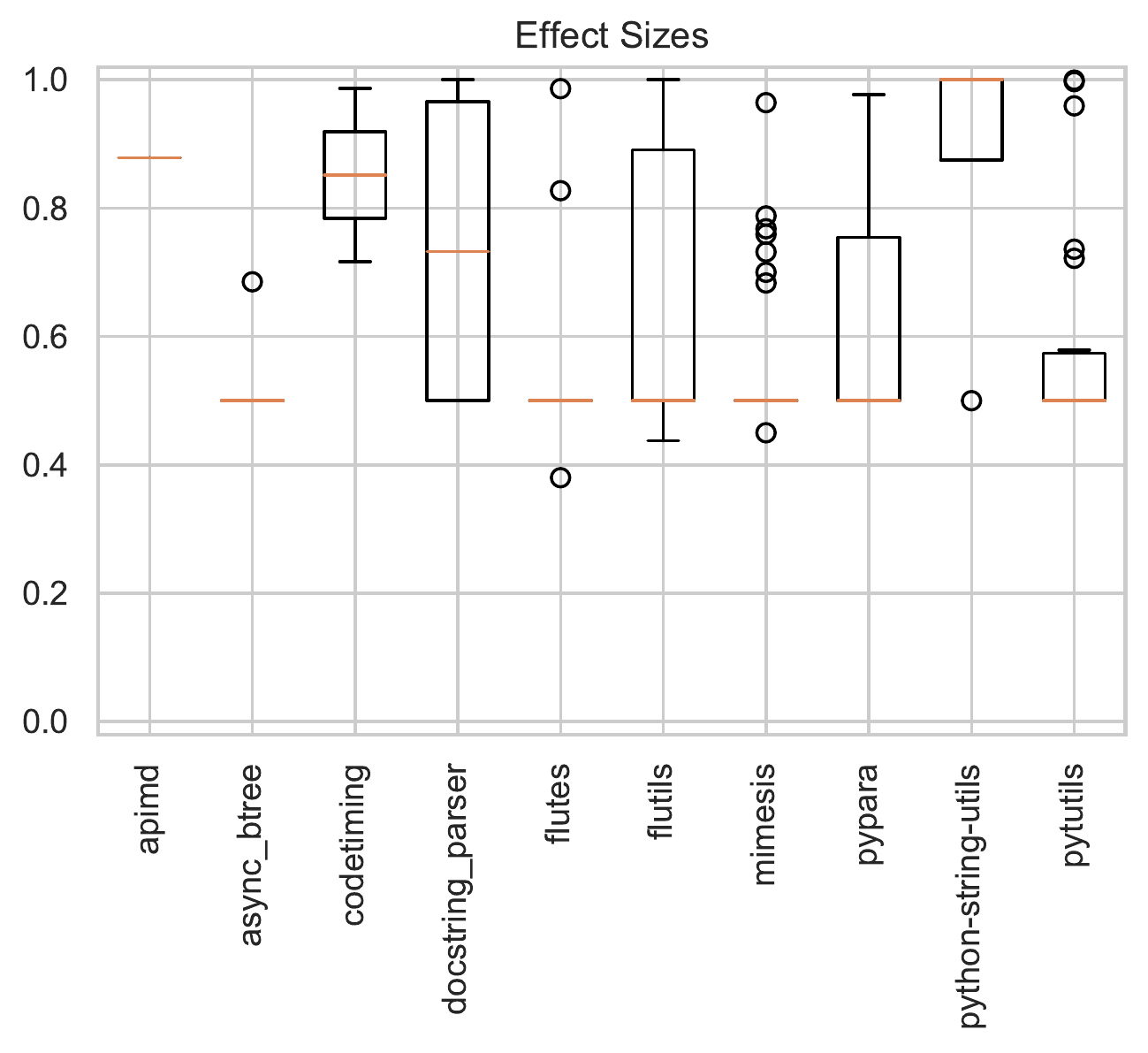}
    \caption{\label{fig:effect-sizes-notypes}\effectsize~effect sizes without
      type information}
  \end{subfigure}
  \caption{\label{fig:effect-sizes}Effect sizes of whole suite versus random
    generation.  Values greater than \num{0.5} indicate whole suite is better
    than random.}
\end{figure}
\begin{figure}[t!]
  \centering
  \includegraphics[width=0.9\textwidth]{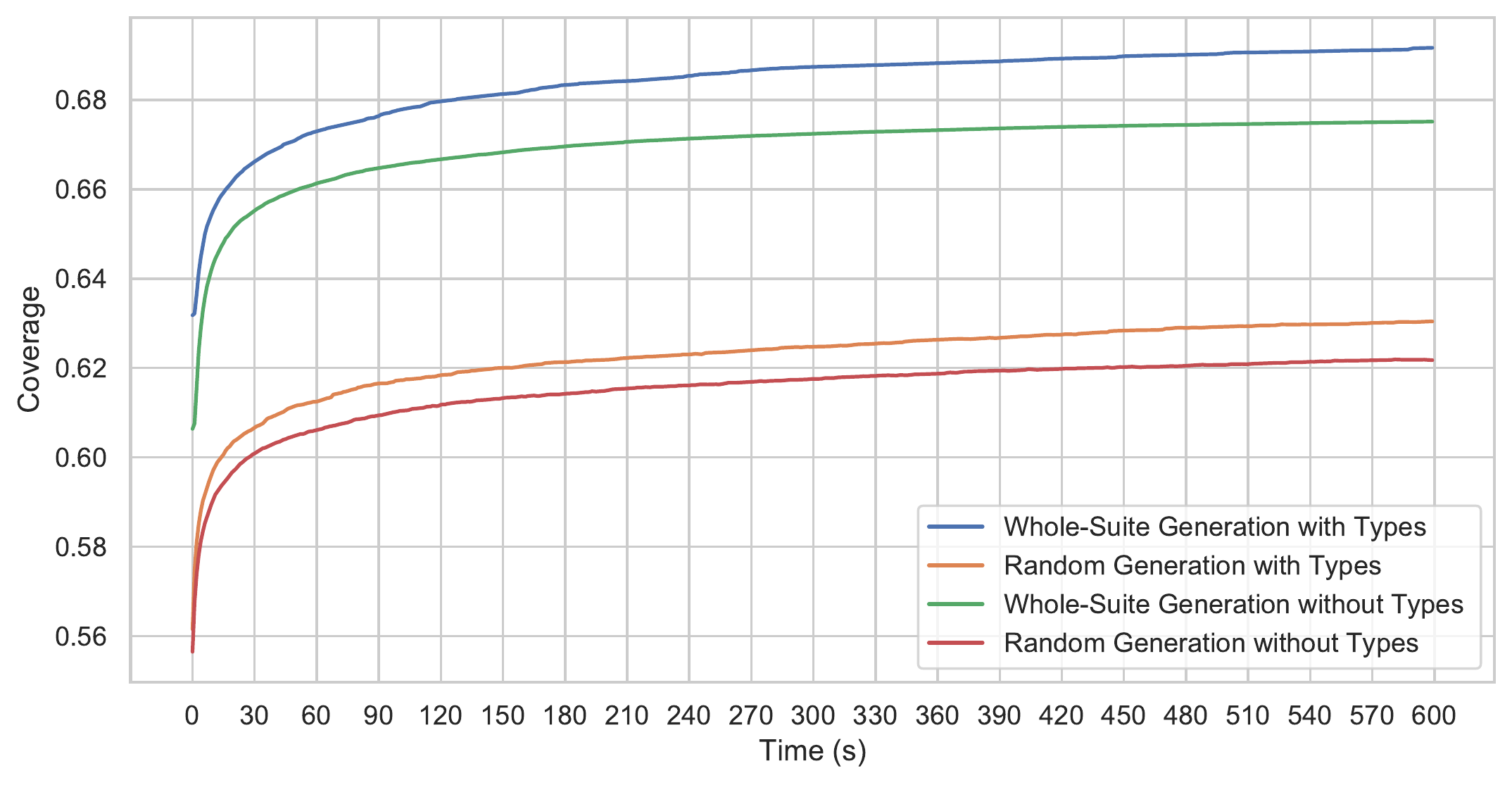}
  \caption{\label{fig:coverage-over-time}Average coverage development over time}
\end{figure}

%
To better understand whether whole-suite test generation performs
better than random test generation,
Fig.~\ref{fig:effect-sizes-typehints} reports the \effectsize{} effect sizes for the per-module comparison of the two
with available type information,
whereas Fig.~\ref{fig:effect-sizes-notypes} reports the same
without available type information.
In both box plots, a value greater than \num{0.5} means that
whole-suite performs better than random test generation,
that is,
yields higher coverage results.
Both plots show that on average whole-suite does not perform worse
than random and, depending on the project,
is able to achieve better results in terms of coverage~(average \effectsize{}
with type information: \num{\AvgEffSizetypehints},
without type information: \num{\AvgEffSizenotypes}).
The effect of these improvements is significant (\(p < 0.05\))
for six out of ten projects,
most notably for
\texttt{apimd}~(\(\effectsize = \num{\EffSizetypehintsapimdA},
\pvalue < 0.001\)),
\texttt{python-string-utils}~(\(\effectsize =
\num{\EffSizetypehintspythonstringutilsA}, \pvalue < 0.001\)),
and \texttt{codetiming}~(\(\effectsize = \num{\EffSizetypehintscodetimingA},
\pvalue = \num{\EffSizetypehintscodetimingPvalue}\)).
For the other projects
the effect is negligible.
In case of \texttt{mimesis}~(\(\effectsize = \num{\EffSizetypehintsmimesisA}\))
this is due to high coverage values
in all configurations---most method parameters expect primitive types,
which are also used for input generation
if no type information is given.
Other projects require specific technical abilities,
for example,
most methods in \texttt{async\_btree}~(\(\effectsize =
\num{\EffSizetypehintsasyncbtreeA}\)) are coroutines,
which require special calls
that cannot currently be generated by \pynguin.
The consequence of this technical limitations is
that \pynguin cannot reach higher coverage
independent of the used algorithm in these cases.
We observed that methods under test often require
collection types as inputs,
in Python prevalently lists and dictionaries.
Also generating these input types
would allow us to execute more parts of the code
which would lead to higher coverage
and thus better results.
We leave this,
however,
as future work.
A further current limitation of our framework
lies in how the available type information is processed.
\pynguin can currently only generate inputs for
concrete types,
unions of types,
and the \mintinline{python}{Any} type---for which it attempts to use
a random type from the pool of available types in the SUT.
Future work shall handle sub-typing relations
as well as generic types~\cite{fraser2014generics}.
Another prevalent parameter type that limits our current tool are callables,
that is,
higher order functions that can be used,
for example,
as call backs.
Previous work has shown
that generating higher-order functions as input types
is feasible for dynamically typed languages
and beneficial for test generation~\cite{selakovic2018higherorderfuncs}.
Furthermore,
\pynguin currently only has a naive static seeding strategy for constants
that incorporates all constant values from the project under test
into test generation,
whereas seeding has been show to have a positive influence
on the quality of test generation~\cite{FA12}
since it allows better-suited input values.
Figure~\ref{fig:coverage-over-time} shows the development
of the average coverage over all modules
over the available generation time of \SI{600}{\second}.
The line plot clearly indicates
that whole-suite generation achieves higher coverage
than random generation,
which again supports our claim.
Overall,
we can answer our first research question as follows:
\summary{RQ\ref{rq:quality}}{%
  Whole-suite test generation achieves
  at least as high coverage as random test generation.
  Depending on the project
  it achieves moderate to strongly higher coverage.
}

\subsection{RQ2: Influence of Type Information}\label{sec:rq2}

\begin{figure}[t!]
  \begin{subfigure}[t]{0.475\textwidth}
    \centering
    \includegraphics[width=\textwidth]{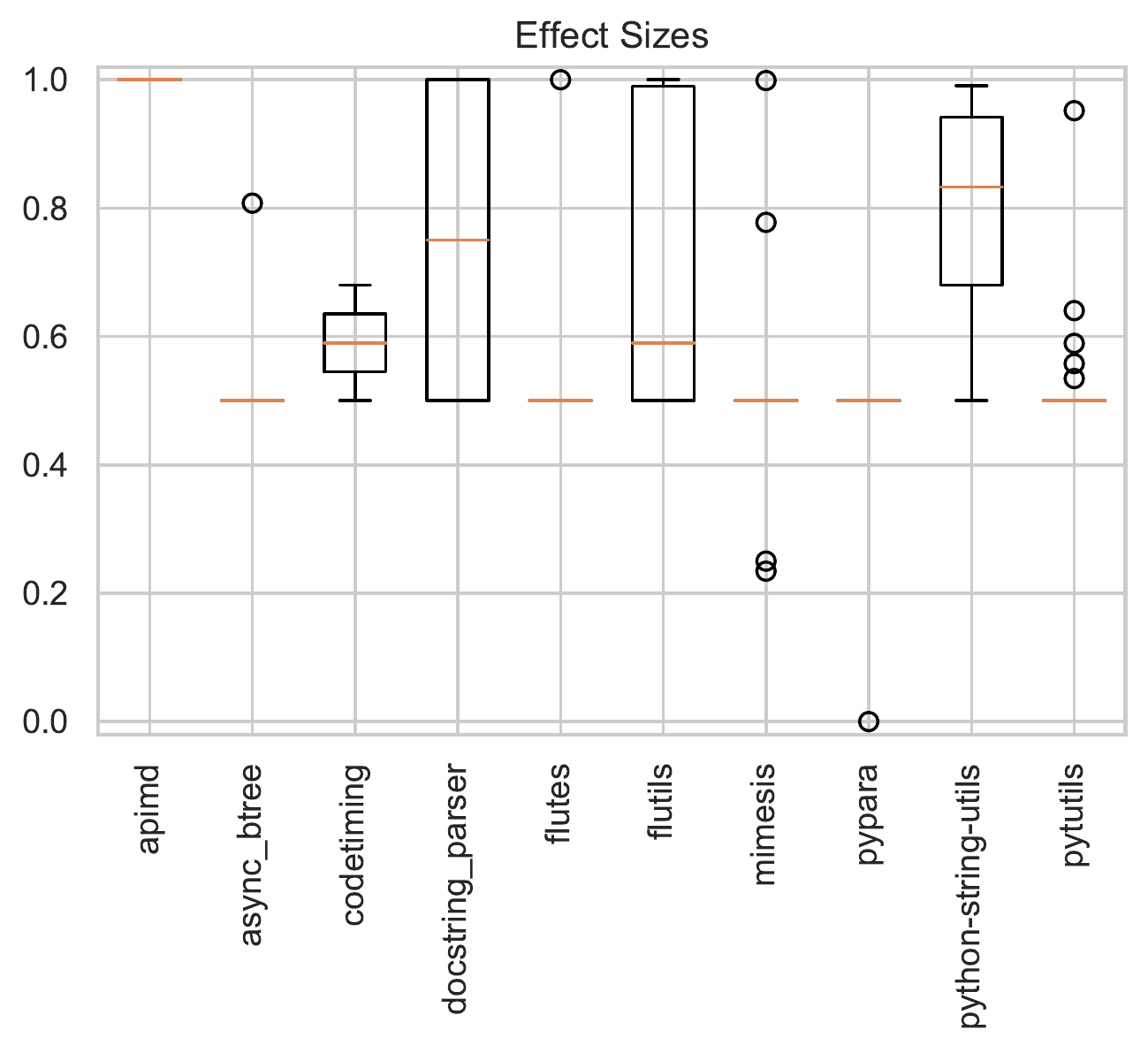}
    \caption{\label{fig:effect-sizes-ws}\effectsize~effect sizes for
      whole-suite generation}
  \end{subfigure}%
  \hfill%
  \begin{subfigure}[t]{0.475\textwidth}
    \centering
    \includegraphics[width=\textwidth]{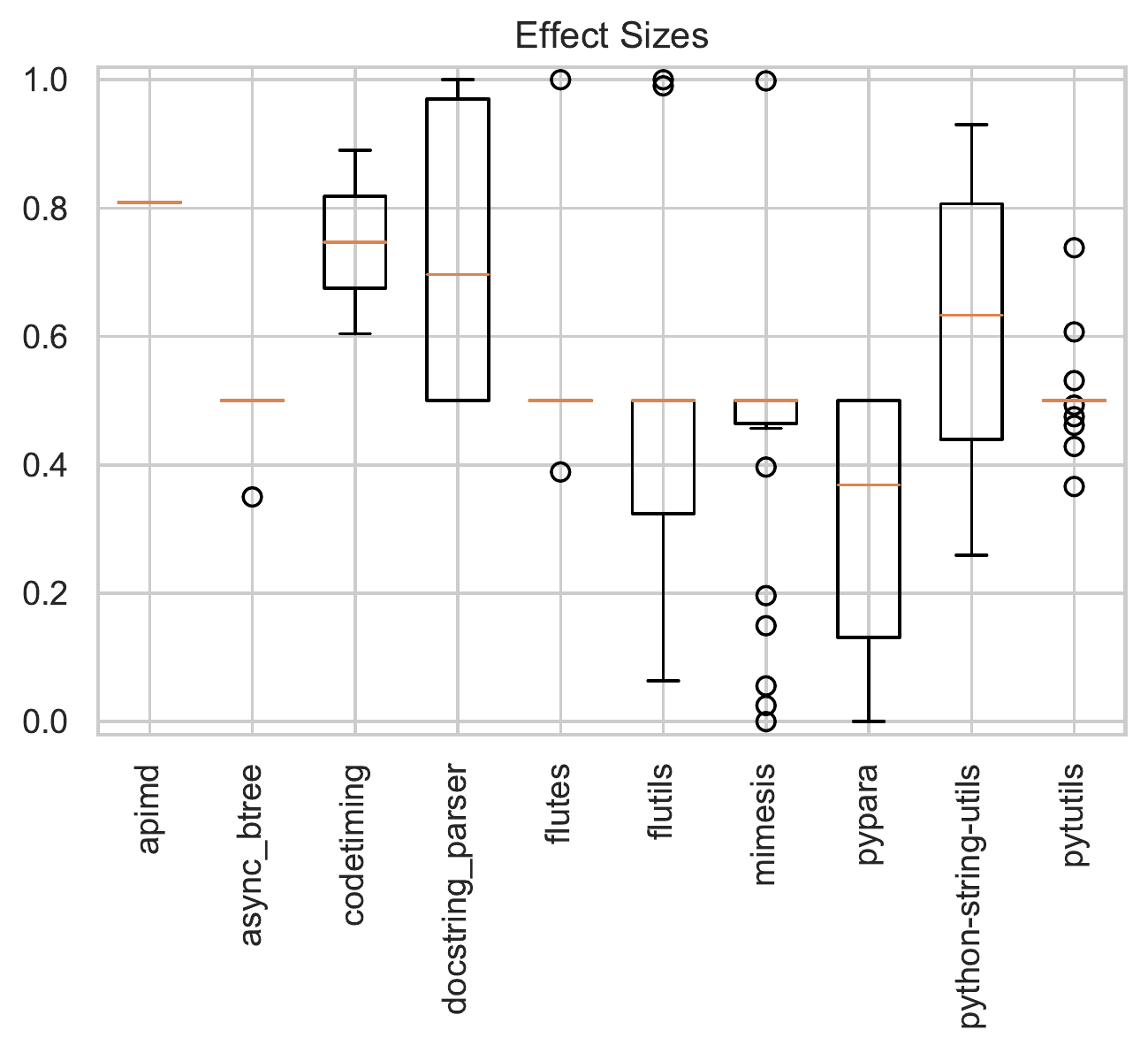}
    \caption{\label{fig:effect-sizes-random}\effectsize~effect sizes for random
      generation}
  \end{subfigure}
  \caption{\label{fig:effect-sizes-type-influence}Effect sizes for type
    influences}
\end{figure}

To answer \mbox{RQ\ref{rq:typeinfluence}}
we compare the coverage values between the configurations with and without type annotations, again using the per-module \effectsize{} effect sizes
on the coverage values.
This time,
we show the effect of type information for whole-suite generation
in Fig.~\ref{fig:effect-sizes-ws}
and for random generation in Fig.~\ref{fig:effect-sizes-random}.
For whole-suite generation,
we observe a large positive effect on some modules, and barely any effect for other modules
when type information is incorporated;
we report an average \effectsize{} value of \num{\AvgEffSizews}
in favour of type information.
For random generation,
we note similar effects,
except for the \texttt{pypara} project~(\(\effectsize =
\num{\EffSizerandompyparaA}, \pvalue = \num{\EffSizerandompyparaPvalue}\));
inspecting the \texttt{pypara} source code reveals
that it uses abstract-class types as type annotations.
\pynguin tries to instantiate the abstract class,
which fails,
and is thus not able to generate method inputs for large parts of the code
because it cannot find an instantiable subtype.
Overall, however,
we report an average \effectsize{} value of \num{\AvgEffSizerandom}
in favour of random generation with type information.


%
The box plots in Fig.~\ref{fig:effect-sizes-type-influence}
indicate similar conclusions for both whole-suite and random testing:
the availability of type information is beneficial for some projects
while its effect is negligible for other projects.
The \texttt{docstring\_parser} project,
for example,
requires their own custom types as parameter values for many methods.
Without type information,
\pynguin has to randomly choose types from all available types,
with a low probability of choosing the correct one,
whereas with available type information
it can directly generate an object of correct type.
Another effect comes in place for the \texttt{python-string-utils} projects:
most of its methods only require primitive input types
but very specific input values.
\pynguin utilises a simple static constant seeding heuristic
for input-value generation.
Due to many values in the constant pool
the chance of picking the correct value is smaller
when not knowing the requested type,
thus leading to lower coverage without type information.
On the other hand,
projects such as \texttt{flutes} require iterables and callables
as parameters in many cases
or need special treatment of their methods
to execute them properly~(see coroutines in \texttt{async\_btree}, for example).
\pynguin currently lacks support to generate these required types,
which prevents larger effects
but does not limit the general approach.
Thus,
the type information cannot be used effectively,
which results in negligible effects between the compared configurations.
%


%
The line plot in Fig~\ref{fig:coverage-over-time} shows the average coverage
per evaluated configuration over the available time for test generation.
It shows that both for whole-suite and random generation
the configuration that incorporates type information
yields higher coverage values
over the full runtime of the generation algorithms,
compared to ignoring type information.
This again supports our claim
that type information is beneficial when generating unit tests
for Python programs.
Overall, we therefore conclude for our second research question:
\summary{RQ\ref{rq:typeinfluence}}{%
  Incorporating type information supports the test generation algorithms
  and allows them to cover larger parts of the code.
  The strength of this effect,
  however,
  largely depends on the SUT.
  Projects that require specific types
  from a large pool of potential types
  benefit more,
  and thus achieve larger effect sizes,
  than projects only utilising simple types.
}

\section{Related Work}\label{sec:related-work}

Closest to our work is  whole-suite test generation in
\toolname{EvoSuite}~\cite{FA13}
and feedback-directed random test generation in
\toolname{Randoop}~\cite{PLE+07}.
Both of these approaches target test generation for Java,
a statically typed language,
whereas our work adapts these approaches
to Python.
To the best of our knowledge, little has been done in the area
of automated test generation for dynamically typed languages.
Approaches such as \toolname{SymJS}~\cite{LAG14}
or \toolname{JSEFT}~\cite{MMP15}
target specific properties of JavaScript web applications,
such as the browser's DOM
or the event system.
Feedback-directed random testing has also been adapted to web applications
with \toolname{Artemis}~\cite{ADJ+2011}.
Recent work proposes
\toolname{LambdaTester}~\cite{selakovic2018higherorderfuncs},
a test generator that specifically addresses the generation
of higher-order functions in dynamic languages.
Our approach,
in contrast,
is not limited to specific application domains.
For automated generation of unit tests for Python we are only aware of
\toolname{Auger}\footnote{%
  \url{https://github.com/laffra/auger}, accessed 2020–07–25.%
};
it generates test cases from recorded SUT executions,
while our approach does the generation automatically.

\section{Conclusions}\label{sec:conclusions}

In this paper we presented \pynguin,
an automated unit test generation framework for Python
that is available as an open source tool, and showed that \pynguin is able to emit unit tests for Python that cover large parts of existing code bases.
\pynguin provides a whole-suite and a random test generation approach,
which we empirically evaluated on ten open source Python projects.
Our results confirm previous findings from the Java world
that a whole-suite approach can outperform a random approach
in terms of coverage.
We further showed that the availability of type information
has an impact on the test generation quality.
Our investigations revealed a range of technical challenges for automated test
generation, which provide ample opportunities for further research,
for example, the integration of further test-generation algorithms, such as
(Dyna)MOSA~\cite{PKT18}, the generation of assertions, or the integration of
type inference approaches.
%
%
%
\bibliographystyle{splncs04}
\bibliography{related}

\begin{thebibliography}{10}
\providecommand{\url}[1]{\texttt{#1}}
\providecommand{\urlprefix}{URL }
\providecommand{\doi}[1]{https://doi.org/#1}

\bibitem{andrews2011genetic}
Andrews, J.H., Menzies, T., Li, F.C.: Genetic algorithms for randomized unit
  testing. {IEEE} Trans. Software Eng.  \textbf{37}(1),  80--94 (2011)

\bibitem{A13}
Arcuri, A.: It really does matter how you normalize the branch distance in
  search-based software testing. Softw. Test. Verification Reliab.
  \textbf{23}(2),  119--147 (2013)

\bibitem{ADJ+2011}
Artzi, S., Dolby, J., Jensen, S.H., M{\o}ller, A., Tip, F.: A framework for
  automated testing of {JavaScript} web applications. In: Proc. ICSE. pp.
  571--580. {ACM} (2011)

\bibitem{baresi2010testful}
Baresi, L., Miraz, M.: Testful: Automatic unit-test generation for java
  classes. In: Proc. ICSE. vol.~2, pp. 281--284. {ACM} (2010)

\bibitem{campos2018empirical}
Campos, J., Ge, Y., Albunian, N., Fraser, G., Eler, M., Arcuri, A.: An
  empirical evaluation of evolutionary algorithms for unit test suite
  generation. Inf. Softw. Technol.  \textbf{104},  207--235 (2018)

\bibitem{csallner2004jcrasher}
Csallner, C., Smaragdakis, Y.: Jcrasher: an automatic robustness tester for
  java. Softw. Pract. Exp.  \textbf{34}(11),  1025--1050 (2004)

\bibitem{FA11}
Fraser, G., Arcuri, A.: Evosuite: Automatic test suite generation for
  object-oriented software. In: Proc. ESEC/FSE. pp. 416--419. {ACM} (2011)

\bibitem{FA12}
Fraser, G., Arcuri, A.: The seed is strong: Seeding strategies in search-based
  software testing. In: Proc. ICST. pp. 121--130. {IEEE} Comp. Soc. (2012)

\bibitem{FA13}
Fraser, G., Arcuri, A.: Whole test suite generation. {IEEE} Trans. Software
  Eng.  \textbf{39}(2),  276--291 (2013)

\bibitem{fraser2014generics}
Fraser, G., Arcuri, A.: Automated test generation for java generics. In: Proc.
  SWQD. LNBPI, vol.~166, pp. 185--198. Springer (2014)

\bibitem{L66}
Levenshtein, V.I.: Binary codes capable of correcting deletions, insertions,
  and reversals. In: Soviet physics doklady. vol.~10, pp. 707--710 (1966)

\bibitem{LAG14}
Li, G., Andreasen, E., Ghosh, I.: {SymJS}: Automatic symbolic testing of
  {JavaScript} web applications. In: Proc. FSE. pp. 449--459. {ACM} (2014)

\bibitem{ma2015grt}
Ma, L., Artho, C., Zhang, C., Sato, H., Gmeiner, J., Ramler, R.: Grt:
  Program-analysis-guided random testing (t). In: Proc. ASE. pp. 212--223.
  {IEEE} Comp. Soc. (2015)

\bibitem{MMP15}
Mirshokraie, S., Mesbah, A., Pattabiraman, K.: {JSEFT}: Automated javascript
  unit test generation. In: Proc. ICST. pp. 1--10. {IEEE} Comp. Soc. (2015)

\bibitem{PLE+07}
Pacheco, C., Lahiri, S.K., Ernst, M.D., Ball, T.: Feedback-directed random test
  generation. In: Proc. ICSE. pp. 75--84. {IEEE} Comp. Soc. (2007)

\bibitem{PKT18}
Panichella, A., Kifetew, F.M., Tonella, P.: Automated test case generation as a
  many-objective optimisation problem with dynamic selection of the targets.
  {IEEE} Trans. Software Eng.  \textbf{44}(2),  122--158 (2018)

\bibitem{sakti2014instance}
Sakti, A., Pesant, G., Gu{\'e}h{\'e}neuc, Y.G.: Instance generator and problem
  representation to improve object oriented code coverage. {IEEE} Trans.
  Software Eng.  \textbf{41}(3),  294--313 (2014)

\bibitem{selakovic2018higherorderfuncs}
Selakovic, M., Pradel, M., Karim, R., Tip, F.: Test generation for higher-order
  functions in dynamic languages. Proc. {ACM} Program. Lang.
  \textbf{2}({OOPSLA}),  161:1--161:27 (2018)

\bibitem{tonella2004evolutionary}
Tonella, P.: Evolutionary testing of classes. In: Proc. ISSTA. pp. 119--128.
  {ACM} (2004)

\end{thebibliography}
\end{document}